\documentclass[aps,prl,floats,twocolumn]{revtex4}

\usepackage[latin1]{inputenc}
\usepackage{graphicx}

\usepackage{amsmath,amsthm,amsfonts,eufrak,amssymb}
\input amssym.def
\newsymbol\square 1003

\newcommand{\p}{P(\theta,\phi)}

\begin{document}

\title{Spin squeezing inequalities and entanglement of $N$ qubit states}
\author{J. K. Korbicz$^1$, J.I. Cirac$^2$, and M. Lewenstein$^{1,3,*}$}
\affiliation{$^1$ Institut f\"ur Theoretische Physik, Universit\"at Hannover, D-30167
Hannover, Germany}
\affiliation{$^2$ Max-Planck Institut f\"ur Quantenoptik, Hans-Kopfermann Str. 1, D-85748, Garching, Germany}

\affiliation{$^3$ ICFO---Institut de Ci\`{e}ncies Fot\`{o}niques, 08034 Barcelona, Spain}

\begin{abstract}
We derive spin squeezing inequalities that generalize the concept
of the spin squeezing parameter and provide necessary and
sufficient conditions for genuine 2-, or 3- qubit entanglement for
symmetric states, and sufficient condition for general $N$-qubit
states. Our inequalities have a clear physical interpretation as
entanglement witnesses, can be easy measured, and are given by
complex, but {\it elementary} expressions.

\end{abstract}

\maketitle

Recently, the
area of quantum correlated systems of atoms or ions, and in particular {\it mesoscopic ionic} and {\it
macroscopic atomic ensembles} \cite{hald} has been developing very rapidly.
Spin squeezing of, say few ions to $10^7$ atoms is
nowadays routinely achieved in such systems. The standard tool to detect the generated forms of
multipartite entanglement \cite{Bouw, primer}
provides the, so called, spin squeezing parameter $\xi^2$ introduced in Ref. \cite{ssparam1}.
The spin squeezing parameter is particularly appreciated by experimentalists for the following reasons: i) it has a clear physical meaning,
ii) it can be relatively easy measured, iii) it
is defined by a simple operation expression, iv) it provides a figure of merit for atomic clocks.  Moreover, as shown in \cite{ssparam2, sanders}, $\xi^2$ is directly connected to entanglement in atomic ensembles, providing a sufficient entanglement condition. However, one should stress that no further investigations  to relate $\xi^2$ to other concepts of quantum information have been carried out so far.

In this Letter we generalize and connect the concept of
spin squeezing parameters to the theory of entanglement witnesses
\cite{witnesses}, i. e. such observables $\mathcal{W}$ that have
non-negative averages for all separable states and there exists
an entangled state $\varrho$ such that
$\text{tr}\big(\varrho\mathcal{W}\big)<0$.
In order to derive the generalized spin squeezing inequalities,
we express state
averages of the appropriate
entanglement witnesses in terms of the macroscopic
spin operators:
\begin{equation}\label{S}
J^i=\sum_{a=1}^N \tfrac{1}{2}\sigma^i_{a}\ \ i=1,2,3 \,\,,
\end{equation}
($\sigma^i$ denote Pauli matrices and indices $a,b,c\dots$
enumerate the particles of the ensemble). We recall
\cite{ssparam1} that a state of a spin-$J$ system is called spin
squeezed if there exists a direction ${\bf n}$, orthogonal to the
mean spin $\langle {\bf J}\rangle$, such that:
\begin{equation}
\xi^2=2\langle \Delta J_{\bf n}^2\rangle/J <1,\label{xi}
\end{equation}
where $J_{\bf n}={\bf n \cdot J}$.

In the proposed approach we begin with considering symmetric
states of $N$ qubits first, i.e. states $\varrho$ supported on the
symmetrized product of individual qubit spaces
$\mathcal{H}_s=\text{Sym}(\mathbb{C}^2\otimes\dots\otimes\mathbb{C}^2)$
($\text{Sym}$ denotes symmetrization). We then use the fact that
for symmetric states of 2, and 3 qubits separability is equivalent
to positivity of the, so called, partial transpose of a state
\cite{eckert} (PPT condition \cite{PPT}). From that we derive the
complete families of generalized spin squeezing inequalities,
which provide necessary and sufficient conditions for genuine 2-,
or 3- qubit entanglement for symmetric states; at the same time
they provide a sufficient condition for general states of $N$
qubits \cite{notka}. Our results imply that spin squeezing leads
to the genuine 2-qubit entanglement (i.e. the corresponding
reduced two-qubit density matrices are entangled) \cite{sanders}.
For symmetric states the converse is also true: 2-qubit entangled
states show a specific type of spin squeezing. In addition, we
obtain somewhat simpler necessary conditions for the 3-qubit case,
that lead to entanglement not implied by the standard spin
squeezing.  The proposed novel inequalities, similarly as the
squeezing parameter, i) have a clear physical meaning in terms of
generalized squeezing and entanglement conditions, ii) can be
relatively easy measured, and iii) are given by complex, but {\it
elementary}  expressions.

The simplest form of entanglement that a multiqubit state $\varrho$ can possess is a 2-qubit entanglement: $\varrho$ is 2-qubit entangled if for some
qubits $a$ and $b$ the reduced density matrix
\begin{equation}
\varrho_{ab}=\text{tr}_{1..\hat{a}..\hat{b}..N}\varrho
\end{equation}
is entangled (the hats over indices mean that those indices are omitted). Let us first consider symmetric states. Then all the reductions $\varrho_{ab}$ are of the same form and act in a symmetric
subspace of $\mathbb{C}^2\otimes\mathbb{C}^2$ - the space of qubits $a$ and $b$.
The PPT criterion \cite{PPT} implies that $\varrho_{ab}$ is
entangled iff there exists a vector $\psi$ such that
\begin{equation}\label{PPT}
\text{tr}_{ab}\big(\varrho_{ab}|\psi\rangle\langle\psi|^{T_1}\big)<0,
\end{equation}
where transpose is defined w.r.t. the standard basis $|0\rangle,|1\rangle$. As $\psi$ we can take any eigenvector of $\varrho_{ab}^{T_1}$, corresponding to a negative eigenvalue.

From the explicit form of $\varrho_{ab}^{T_1}$ we deduce that
$|\psi\rangle$ can be parametrized as follows \cite{symm}:
$|\psi\rangle=\eta |00\rangle+\beta
|01\rangle+\beta^*|10\rangle+\gamma |11\rangle$, with $\alpha,
\gamma\in \mathbb{R}$. Hence the coefficients of $|\psi\rangle$
form a hermitean matrix: $[\psi_{CD}]_{C,D=0,1}$. We can
diagonalize it:
$\psi_{CD}=\tilde{U}_{AC}^*\Delta_{AB}\tilde{U}_{BD}$, where
$\Delta=\text{diag}(\text{sin}\tfrac{\alpha}{2},\pm
\text{cos}\tfrac{\alpha}{2})$, $-\pi\le \alpha \le \pi$,
$\tilde{U}\in SU(2)$, and then define
$U=\sum_{C,D=0,1}\tilde{U}_{CD}|D\rangle\langle C|$ to finally
obtain the following parametrization:
\begin{equation}\label{psi2}
|\psi\rangle=U^*\otimes U |\psi_0\rangle\,,\ \ |\psi_0\rangle=\text{sin}\tfrac{\alpha}{2} |00\rangle+\text{cos}\tfrac{\alpha}{2}|11\rangle\,,
\end{equation}
(we have fixed the overall phase). Substituting (\ref{psi2}) into (\ref{PPT}) leads to the condition:
\begin{equation}\label{PPT2}
\text{tr}_{ab}\big(\varrho_{ab}U \otimes U |\psi_0\rangle\langle\psi_0|^{T_1}U^\dagger \otimes U^\dagger\big)<0\,.
\end{equation}

Note, that $|\psi_0\rangle\langle\psi_0|^{T_1}$ can be decomposed into Pauli matrices:
\begin{eqnarray}\label{gh}
& & |\psi_0\rangle\langle\psi_0|^{T_1}=\tfrac{1}{4} \,\text{sin}^2\tfrac{\alpha}{2} \big({\bf 1}+\sigma^z\big)\otimes \big({\bf 1}+\sigma^z\big)\nonumber\\
& & +\tfrac{1}{4} \,\text{cos}^2\tfrac{\alpha}{2} \big({\bf 1}-\sigma^z\big)\otimes\big({\bf 1}-\sigma^z\big)\\
& & +\tfrac{1}{4}\,\text{sin}\alpha \big(\sigma^x\otimes\sigma^x+\sigma^y\otimes\sigma^y\big)\,,\nonumber
\end{eqnarray}
and the adjoint action of $SU(2)$ in (\ref{PPT2}) induces a $SO(3)$ rotation $R$ of $\sigma^i$: $U \sigma^i U^\dagger=R^i_{\phantom{i}j}\sigma^j$ (here and throughout we sum over repeated indices). We will denote the axes of the rotated frame by $\bf{k},\bf{l},\bf{n}$.

Using (\ref{gh}) we can express the inequality (\ref{PPT2})
through the rotated total spin operators (\ref{S}). We first
observe that
$\text{tr}_{ab}\big(\varrho_{ab}|\psi\rangle\langle\psi|^{T_1}\big)=\text{tr}\big(\varrho
|\psi_{ab}\rangle\langle\psi_{ab}|^{T_1}\big)$, where
$|\psi_{ab}\rangle$ is the natural embedding of $|\psi\rangle$
into $\mathcal{H}$. Since all $\varrho_{ab}$ are of the same form,
we can sum (\ref{PPT2}) over all pairs of qubits: $\sum_{\langle
ab\rangle}=\sum_{a=1}^{N-1}\sum_{b=a+1}^{N}$ and use the identity:
$\sum_{\langle ab\rangle}\sigma ^i_a\otimes\sigma ^i_b=2
(J^i)^2-N/2$ to obtain the following inequality \cite{spiny}:
\begin{eqnarray}
& & \text{sin}\alpha\big(\tfrac{N^2}{4}-\langle J^2_{\bf n}\rangle\big)-(N-1)\text{cos}\alpha \langle J_{\bf n} \rangle\nonumber\\
& &+\langle J^2_{\bf n}\rangle +\tfrac{N(N-2)}{4}<0\,,\label{2x2}
\end{eqnarray}
where the averages are taken w.r.t. $\varrho$.

Let us now fix the direction ${\bf n}$ and minimize the l.h.s. of the inequality (\ref{2x2}) w.r.t. $\alpha$. We find that the inequality (\ref{2x2}) is satisfied if and only if:
\begin{equation}\label{2x2min}
\langle J^2_{\bf n}\rangle +\tfrac{N(N-2)}{4}<\sqrt{\big(\tfrac{N^2}{4}-\langle J^2_{\bf n}\rangle\big)^2+(N-1)^2\langle J_{\bf n} \rangle^2}\,.
\end{equation}

For a general, i.e. not necessarily symmetric, state $\varrho$ we
can still test entanglement of all the bipartite reductions
$\varrho_{ab}$ with the {\it same} vector (\ref{psi2}). The sum
$2\text{tr}\big(\varrho \sum_{\langle ab\rangle}
|\psi_{ab}\rangle\langle\psi_{ab}|^{T_1}\big)$ is then not greater
than the l.h.s of (\ref{2x2}) due to \cite{spiny} and we finally
obtain from (\ref{2x2min}):

{\bf Criterion for bipartite entanglement.} {\it If there exists a direction $\bf{n}$ such that the following inequality holds:
\begin{equation}\label{2x2'}
\frac {4 \langle \Delta J_{\bf n}^2\rangle}{N}< 1- \frac{4\langle J_{\bf n} \rangle^2}{N^2}
\end{equation}
then the state $\varrho$ possesses bipartite entanglement. For symmetric states the above condition is both necessary and sufficient.}

To relate the above criterion to the standard spin squeezing condition (\ref{xi}) for spin-$J$ states, note that if (\ref{xi}) is
satisfied for some direction ${\bf n}$ then so is (\ref{2x2'}) as $\langle J_{\bf n} \rangle=0$ and $J\le N/2$. Hence, spin squeezed
states possess 2-qubit entanglement (for symmetric states this was proven in Ref. \cite{sanders}). For symmetric states,
 for which $J=N/4$, the (modified) converse also holds: condition (\ref{2x2'}) implies existence of a spin component $J_{\bf n}$ such
 that $\langle \Delta J_{\bf n}^2\rangle<N/4$. This differs from the standard definition of spin squeezing (\ref{xi}) in that the
 direction ${\bf n}$ need not be orthogonal to $\langle {\bf J} \rangle$. Nevertheless, we also call such states spin squeezed.

Let us now consider the case when $\varrho$ possesses genuine 3-qubit entanglement,
i.e. for some triple of qubits $abc$, the reduced
density matrix
\begin{equation}
\varrho_{abc}=\text{tr}_{1..\hat{a}..
\hat{b}..\hat{c}..N}\varrho
\end{equation}
is 3-party entangled. If we again consider symmetric states first,
then  PPT criterion is still necessary and sufficient for
separability, since
$\text{Sym}(\mathbb{C}^2\otimes\mathbb{C}^2\otimes\mathbb{C}^2)$
is a subspace of $\mathbb{C}^2\otimes\mathbb{C}^3$. Thus we can
proceed as before.

A vector $|\psi\rangle$, corresponding to any negative eigenvalue of $\varrho_{abc}^{T_1}$ must be necessarily a 3-party entangled vector from $\mathbb{C}^2\otimes\text{Sym}(\mathbb{C}^2\otimes \mathbb{C}^2)$. The parametrization of such vectors was found in Ref. \cite{Dur}; there are two families:
\begin{eqnarray}
& & |\psi\rangle = A \otimes B \otimes B |GHZ\rangle  \label{GHZ}\\
& & |\psi\rangle = A \otimes U \otimes U |W\rangle \label{W} \,,
\end{eqnarray}
where matrices $A,B \in SL(2,\mathbb{C})$, $U\in SU(2)$, and
$|GHZ\rangle=(1/\sqrt{2})(|000\rangle+|111\rangle)$,
$|W\rangle=(1/\sqrt{3})(|001\rangle+|010\rangle+|100\rangle)$. The
action of $SL(2,\mathbb{C})$ on the Pauli matrices in the
decomposition of $|\psi\rangle\langle\psi|^{T_1}$ now induces
restricted, i.e. orientation and time-orientation preserving,
Lorenz transformations:
\begin{equation}
A^*\sigma^\mu A^T=\Lambda^\mu_{\phantom{\mu}\nu}\sigma^\nu\,, \ B\sigma^\mu B^\dagger=L^\mu_{\phantom{\mu}\nu}\sigma^\nu\,,\ \sigma^0={\bf 1}\,,
\end{equation}
(Greek indices run through $0\dots 4$). Hence we obtain the following inequality, analogous to (\ref{PPT2}):
\begin{equation}\label{poly2}
\text{tr}_{abc}\big(\varrho_{abc}|\psi\rangle\langle\psi|^{T_1}\big)=\tfrac {1}{8} K_{\alpha\beta\gamma} \langle \sigma_a^{\alpha} \otimes \sigma_b^\beta \otimes \sigma_c^{\gamma} \rangle < 0 \,,
\end{equation}
where
\begin{eqnarray}
& & K_{\alpha\beta\gamma}(\Lambda,L,L)\!= \!\Lambda^0_{\phantom{\mu}\alpha} L^0_{\phantom{\mu}\beta}L^0_{\phantom{\mu}\gamma}\!\!+\!\Lambda^0_{\phantom{\mu}\alpha} L^3_{\phantom{\mu}\beta}L^3_{\phantom{\mu}\gamma}\!\!+\!\Lambda^1_{\phantom{\mu}\alpha} L^1_{\phantom{\mu}\beta} L^1_{\phantom{\mu}\gamma}\nonumber\\
& & + 2 \Lambda^3_{\phantom{\mu}\alpha} L^0_{\phantom{\mu}(\beta} L^3_{\phantom{\mu}\gamma)}-\Lambda^1_{\phantom{\mu}\alpha} L^2_{\phantom{\mu}\beta} L^2_{\phantom{\mu}\gamma}+ 2\Lambda^2_{\phantom{\mu}\alpha} L^1_{\phantom{\mu}(\beta} L^2_{\phantom{\mu}\gamma)}\,,\label{KGHZ}
\end{eqnarray}
for the $GHZ$ family (\ref{GHZ}), or
\begin{eqnarray}
& & K_{\alpha\beta\gamma}(\Lambda,R,R)= \tfrac{1}{3}\Big\{3\Lambda^0_{\phantom{\mu}\alpha} R^0_{\phantom{\mu}\beta}R^0_{\phantom{\mu}\gamma}-\!3\Lambda^3_{\phantom{\mu}\alpha} R^3_{\phantom{\mu}\beta}R^3_{\phantom{\mu}\gamma}\nonumber\\
& & \!+ 2\Lambda^0_{\phantom{\mu}\alpha} R^0_{\phantom{\mu}(\beta}R^3_{\phantom{\mu}\gamma)}\!+\!\Lambda^3_{\phantom{\mu}\alpha} R^0_{\phantom{\mu}\beta}R^0_{\phantom{\mu}\gamma}\!-\!\Lambda^0_{\phantom{\mu}\alpha} R^3_{\phantom{\mu}\beta}R^3_{\phantom{\mu}\gamma}\label{KW}\\
& & \!-2\Lambda^3_{\phantom{\mu}\alpha} R^0_{\phantom{\mu}(\beta}R^3_{\phantom{\mu}\gamma)}\!+\!4\Lambda^1_{\phantom{\mu}\alpha} R^0_{\phantom{\mu}(\beta}R^1_{\phantom{\mu}\gamma)}\!+\!4\Lambda^1_{\phantom{\mu}\alpha} R^1_{\phantom{\mu}(\beta}R^3_{\phantom{\mu}\gamma)}\nonumber\\
& & -4\Lambda^2_{\phantom{\mu}\alpha} R^0_{\phantom{\mu}(\beta}R^2_{\phantom{\mu}\gamma)}-4\Lambda^2_{\phantom{\mu}\alpha} R^2_{\phantom{\mu}(\beta}R^3_{\phantom{\mu}\gamma)}\Big\}\,\nonumber
\end{eqnarray}
for the $W$ family (\ref{W}). Here $R^{\mu}_{\phantom{\mu}\nu}$ is the four-dimensional embedding of the rotation generated by $U$ from (\ref{W}) and round brackets denote symmetrization.

In order to express the inequality (\ref{poly2}) through
$\varrho$-averages of the spin operators (\ref{S}), we introduce
an artificial time-component $J^0=(N/2){\bf 1}$. The operators
$J^\mu=(J^0,J^i)$ do not constitute relativistic generalization of
the operators $J^i$ and we introduce them just for notational
reasons. Since $\varrho_{abc}$ is symmetric, the indices
$\alpha\beta\gamma$ in (\ref{poly2}) can be symmetrized. Then
after summing (\ref{poly2}) over all triples of qubits
$\sum_{\langle abc\rangle}=
\sum_{a=1}^{N-2}\sum_{b=a+1}^{N-1}\sum_{c=b+1}^N$, we can use the
identity:
\begin{eqnarray}
& & 3 \sum_{\langle abc\rangle} \sigma_a^{(\alpha}\otimes\sigma_b^\beta\otimes\sigma_c^{\gamma)}=4 J^{(\alpha}J^\beta J^{\gamma)}-6 f^{(\alpha\beta}_{\phantom{(\alpha\beta}\mu} J^{\gamma}J^{\mu)}\nonumber\\
& & + 2 f^{(\alpha\beta}_{\phantom{(\alpha\beta}\mu}f^{\gamma\mu)}_{\phantom{(\alpha\beta}\nu}J^\nu\,,\label{masakra}
\end{eqnarray}
where $f^{0\alpha}_{\phantom{0\alpha}\beta}=f^{\alpha 0}_{\phantom{0\alpha}\beta}=\delta^\alpha_{\phantom{\alpha}\beta}$, $f^{ij}_{\phantom{jk}\alpha}=\text{i}\sum_l\epsilon^{ijl}\delta^l_{\phantom{l}\alpha}+\delta^{ij}\delta^0_{\phantom{0}\alpha}$, to finally obtain:

{\bf Criterion for tripartite entanglement.} {\it A symmetric state $\varrho$ possesses a genuine tripartite entanglement iff there exist two restricted Lorenz transformations $\Lambda$, $L$, or a restricted Lorenz transformation $\Lambda$ and a rotation $R$, such that:
\begin{equation}
K_{(\alpha\beta\gamma)}\Big\{2\langle J^{\alpha}J^\beta J^{\gamma}\rangle -3f^{\alpha\beta}_{\phantom{\alpha\beta}\mu}\langle J^{(\gamma}J^{\mu)}\rangle+f^{\alpha\beta}_{\phantom{\alpha\beta}\mu}f^{(\gamma\mu)}_{\phantom{(\alpha\beta)}\nu}\langle J^\nu\rangle\Big\}<0\label{2x2x2}
\end{equation}
holds, with $K_{\alpha\beta\gamma}$ given by (\ref{KGHZ}), or by (\ref{KW}) respectively.}

The above criterion serves also as a sufficient condition for tripartite entanglement for a general state $\varrho$, with the modification that $K(\Lambda,L,L)$ or $K(\Lambda,R,R)$ in (\ref{2x2x2}) have to be substituted with $1/3\big[K(\Lambda,L,L)+K(L,\Lambda,L)+K(L,L,\Lambda)\big]$ or $1/3\big[K(\Lambda,R,R)+K(R,\Lambda,R)+K(R,R,\Lambda)\big]$ respectively to achieve the index symmetrization.

The search for matrices $\Lambda$, $L$ can be difficult due to non-compactness of the restricted Lorenz group. It is therefore desirable to develop some simpler conditions as well. For mesoscopic systems with not too large $N$ we may do so, using some specific witnesses that detect genuine $GHZ$-type, or genuine $W$-type entanglement, found in Ref. \cite{3qbit}:
\begin{eqnarray}
& &\mathcal{W}_{GHZ}=\tfrac{3}{4}{\bf 1}-|GHZ\rangle\langle GHZ|\label{wghz}\\
& &\mathcal{W}_{W_1}=\tfrac{2}{3}{\bf 1}-|W\rangle\langle W|\label{ww1}\\
& &\mathcal{W}_{W_2}= \tfrac{1}{2}{\bf 1}-|GHZ\rangle\langle GHZ|\label{ww2}\,,
\end{eqnarray}
where, in order to be more general, we may now define the vectors
$|GHZ\rangle$ and $|W\rangle$ in an arbitrary frame
$\bf{k},\bf{l},\bf{n}$, rotated w.r.t. the original one. The
witnesses $\mathcal{W}_{GHZ}$  detects states of GHZ class which
are neither of the W class, nor biseparable. Finally, the
witnesses $\mathcal{W}_{W_1}$ and $\mathcal{W}_{W_2}$ detect
states of GHZ- or W-class, which are not biseparable \cite{3qbit}.
Proceeding as before and using the same witnesses
(\ref{wghz})-(\ref{ww2}) for all tripartite reductions
$\varrho_{abc}$ of a general state $\varrho$, we get necessary
conditions for:

{\bf GHZ-type entanglement.} {\it If for a state $\varrho$ there exist orthogonal directions ${\bf k,l,n}$
such that the following inequality is fulfilled
\begin{eqnarray}
& & -\tfrac{1}{3}\langle J_{\bf k}^3\rangle+\langle J_{\bf l}J_{\bf k}J_{\bf l}\rangle-\tfrac{N-2}{2}\langle J_{\bf n}^2\rangle+\tfrac{1}{3}\langle J_{\bf k}\rangle\nonumber\\
& & +\tfrac{N(N-1)(5N-2)}{24}<0\label{ss1}\,,
\end{eqnarray}
then the  state $\varrho$ possesses a genuine GHZ-type entanglement.}

{\bf GHZ- or W-type entanglement.} {\it If for a state $\varrho$ there exist orthogonal directions ${\bf k,l,n}$
such that one of the following inequalities is fulfilled
\begin{eqnarray}
& & \langle J_{\bf n}^3\rangle-2\langle J_{\bf l}J_{\bf n}J_{\bf
l}\rangle-2\langle J_{\bf k}J_{\bf n}J_{\bf k}\rangle\nonumber\\
& & -\tfrac{N-2}{2}\big(2\langle J_{\bf k}^2\rangle+2\langle J_{\bf
l}^2\rangle-\langle J_{\bf
n}^2\rangle\big)-\tfrac{N^2-4N+8}{4}\langle
J_{\bf n}\rangle\nonumber\\
& & +\tfrac{N(N-2)(13N-4)}{24}<0\label{ss2}\\
 & &-\tfrac{1}{3}\langle J_{\bf k}^3\rangle+\langle J_{\bf l}J_{\bf k}J_{\bf l}\rangle-\tfrac{N-2}{2}\langle J_{\bf n}^2\rangle+\tfrac{1}{3}\langle J_{\bf k}\rangle\nonumber\\
& &+\tfrac{N^2(N-2)}{8}<0\label{ss3}\,,
\end{eqnarray}
then the  state $\varrho$ possesses a genuine 3-qubit (GHZ- or W-type) entanglement.}

The above spin squeezing criteria (\ref{2x2'}),(\ref{2x2x2}), (\ref{ss1})-(\ref{ss3}) constitute the main result of this Letter.
The inequalities (\ref{ss1}) and (\ref{ss3}) can be further simplified if we choose the directions ${\bf k,n}$ such that  $\langle J_{\bf k}\rangle=\langle J_{\bf n}\rangle=0$. Let us further assume that: $\langle J_{\bf k}^2\rangle\ge N/4$, $\langle J_{\bf n}^2\rangle\ge N/4$, so that there is no spin-squeezing in the sense of the definition (\ref{xi}). Then from criteria (\ref{ss1}), (\ref{ss3}) it follows, that if:
\begin{equation}
-\tfrac{1}{3}\langle J_{\bf k}^3\rangle+\langle J_{\bf l}J_{\bf k}J_{\bf l}\rangle+\tfrac{N(5N^2-10N+8)}{24}<0\label{ss1'}
\end{equation}
or:
\begin{equation}
-\tfrac{1}{3}\langle J_{\bf k}^3\rangle+\langle J_{\bf l}J_{\bf k}J_{\bf l}\rangle+\tfrac{N(N-1)(N-2)}{8}<0\label{ss2'}
\end{equation}
holds, then the state $\varrho$ possesses a genuine $GHZ$ or 3-qubit entanglement respectively. Thus, in this specific situation, the inequalities (\ref{ss1'}) and (\ref{ss2'}) detect a different type of entanglement than that implied by the standard spin squeezing \cite{ssparam2, sanders}.

Generalization of the above procedure to study the entanglement between more qubits
is straightforward - one uses inequalities of the type $\text{tr}\big(\varrho\mathcal{W}\big)<0$ with
appropriable witnesses $\mathcal{W}$. However, for the case of four or
more qubits the PPT criterion is no longer sufficient and only necessary conditions of the type (\ref{ss1})-(\ref{ss3}) can be obtained.

Let us conclude with a general remark concerning full (i.e. N-qubit) separability of a
symmetric state and a connection to the method of Ref. \cite{my}. Every symmetric state $\varrho$ of $N$-qubits admits an analog of Glauber-Sudarshan $P$-representation \cite{glauber,perelomov}:
\begin{equation}\label{p-rep}
\varrho=\int_{\mathbb{S}^2} d\Omega P(\theta,\phi)\, |\theta,\phi\rangle\langle\theta,\phi|\otimes\dots\otimes|\theta,\phi\rangle\langle\theta,\phi|\,,
\end{equation}
where $d\Omega=\text{sin}\theta\,d\theta d\phi$ is the volume element on the Bloch sphere, and $|\theta,\phi\rangle=\text{cos}(\tfrac{\theta}{2})\,|0\rangle+\text{e}^{\text{i}\phi}\text{sin}(\tfrac{\theta}{2})\,|1\rangle$ is a  spin coherent state of a single qubit. Note that every qubit is representable in this form. The representation (\ref{p-rep}) is not unique, as in the decomposition of $\p$ over spherical harmonics $Y_{lm}$, $\varrho$ determines only terms with $l \le N$, and hence $\p$ can be chosen to be a polynomial in the Cartesian coordinates on the sphere. Now the following fact holds \cite{braunstein, Kraus}:

{\it A symmetric state $\varrho$ is fully separable iff there
exists a representation (\ref{p-rep}) where $\p d\Omega$ is an
element of a probabilistic measure on $\mathbb{S}^2$}

{\it Proof}. Implication $\Leftarrow$ is obvious as the integral
in (\ref{p-rep}) is a norm limit of separable states. To prove the
implication $\Rightarrow$, observe that if $\varrho$ is separable,
then it can be decomposed as $\varrho=\sum_{k}p_k
|\theta_k,\phi_k\rangle\langle\theta_k,\phi_k|\otimes\cdots\otimes
|\theta_k,\phi_k\rangle\langle\theta_k,\phi_k|$, $p_k\ge 0$, $\sum
p_k=1$, as vectors of the form
$|\theta_k,\phi_k\rangle\langle\theta_k,\phi_k|\otimes\cdots\otimes
|\theta_k,\phi_k\rangle\langle\theta_k,\phi_k|$ are the only
symmetric product vectors. We define then $\p=\sum_k p_k
\delta(\text{cos}\theta-\text{cos}\theta_k)\delta(\phi-\phi_k)$;
the expansion of $\delta$'s over $Y_{lm}$ can be truncated at
$l=N$ $\Box$

We observe that if $\mathcal{W}$ is an entanglement witness, then:
\begin{equation}
\text{tr}\big(\varrho\mathcal{W}\big)=\int\! d\Omega \p \, \, w(\theta,\phi)
\end{equation}
where $w(\theta,\phi)=\langle(\theta,\phi)^{\otimes
N}|\mathcal{W}|(\theta,\phi)^{\otimes N}\rangle$ is a positive
semidefinite polynomial of the $N$th order in the Cartesian
coordinates. Hence, the criteria (\ref{2x2}) and (\ref{2x2x2}),
with the reversed inequality signs, can be interpreted as
necessary and sufficient conditions for $\p d\Omega$ to be an
element of a probabilistic measure for $N=2,3$ respectively.

The above fact establishes an interesting link between separability of symmetric states and the problem of description of classical states of a $1D$ harmonic oscillator \cite{vogel, my}. In the latter problem, classical states are in one-to-one correspondence with probabilistic measures on $\mathbb{R}^2$. We have proved in \cite{my} that, among some specific subclass of states, the classical ones are detected by observables, arising from positive semidefinite (psd) polynomials which are sums of squares of other polynomials.

Summarizing, we have introduced a method of deriving generalized
spin squeezing inequalities, that characterize genuine $N$-qubit
entanglement. The results of the paper provide connection of spin
squeezing to entanglement witnesses, and an alternative physical
meaning to spin squeezing as qualitative and quantitative
characterization of the $N$-qubit entanglement. The inequalities
can be directly measured and provide novel entanglement detection
tools for macroscopic atomic ensembles.

We thank A. Ac\'\i n, M. Mitchell and J. Eschner for discussions, and  the  Deutsche Forschungsgemeinschaft
(SFB 407, SPP 1078, GK 282, 436 POL) and the EU Programme QUPRODIS for support.


\begin{thebibliography}{999}

\bibitem[*] aAlso at Instituci\'o Catalana de Recerca i Estudis Avan\c cats.

\bibitem{hald} For pioneering experimental work, see J. Hald, J.L. S\o rensen,
C. Schori, and E.S. Polzik, Phys. Rev. Lett. {\bf 83}, 1319 (1999);
for very recent experiments on quantum feedback control, see
J.M. Geremia, J.K. Stockton, and H. Mabuchi, Science {\bf 304}, 270 (2004);
for the recent review of spin squeezing
using Gaussian states, see L. B. Madsen and K. M\o lmer,
Phys. Rev. A{\bf 70}, 052324 (2004).

\bibitem{Bouw} D. Bouwmeester, A. Ekert, and A. Zeilinger (eds.), {\it The physics of quantum information} (Springer, Berlin, 2000).

\bibitem{primer} M. Lewenstein {\it et al.,} J. Mod. Opt., {\bf 47}, 2481 (2000).

\bibitem{ssparam1} M. Kitagawa and M. Ueda, Phys. Rev. A, {\bf 47}, 5138 (1993).

\bibitem{ssparam2} A. S\o rensen, L. M. Duan, J. I. Cirac,
and P. Zoller, Nature {\bf 409}, 63 (2001).

\bibitem{sanders} X. Wang and B. C. Sanders, Phys. Rev. A {\bf 68}, 012101 (2003)

\bibitem{witnesses} M. Horodecki, P. Horodecki, and R. Horodecki, Phys. Lett. A {\bf 223}, 1 (1996); B.M. Terhal, Phys. Lett. A {\bf 271}, 319 (2000); M. Lewenstein, B. Kraus, J. I. Cirac, and P. Horodecki, Phys. Rev. A {\bf 62}, 052310 (2000).

\bibitem{eckert} K. Eckert, J. Schliemann, D. Bru\ss, and M. Lewenstein, Ann. Phys. {\bf 299}, 88 (2002).

\bibitem{PPT} A.Peres, Phys. Rev. Lett. {\bf 77}, 1413 (1996).



\bibitem{notka} Note, that macroscopic atomic ensembles
may be prepared in symmetric subspace, although
ultimately individual spontaneous emission acts take
the system out of this subspace. Nevertheless, the
symmetric component might remain significant for long times.


\bibitem{symm} For a general bipartite symmetric state we have:
$$\varrho^{T_1}=\left [ \begin{array}{cccc} \epsilon_0 & \delta & \delta^* & \tau\\
                                      \delta^* & \epsilon_1 & \omega^* & \pi^* \\
                      \delta & \omega & \epsilon_1 & \pi\\
                      \tau & \pi & \pi^* & \epsilon_2 \end{array}\right]$$
with $\epsilon_0,\epsilon_1,\epsilon_2$ and $\tau$ real. Then it
is easy to chek that vectors of the type $\eta |00\rangle+\beta
|01\rangle+\beta^*|10\rangle+\gamma |11\rangle$ are preserved by
$\varrho^{T_1}$ and since they have three independent parameters
(we take them to be normalized) it is possible to find a solution
of the eigenvalue equation.

\bibitem{spiny} Note that generally $\langle J_{\bf k}^2\rangle +\langle J_{\bf l}^2\rangle +\langle J_{\bf n}^2\rangle \le N(N+2)/4$ and for symmetric states the equality holds.


\bibitem{Dur} W. D\"ur, G. Vidal, and J.I. Cirac, Phys. Rev A {\bf 62}, 062314 (2000).


\bibitem{3qbit} A. Ac\'\i n, D. Bru\ss, M. Lewenstein and A. Sanpera, Phys. Rev. Lett. {\bf 87}, 040401 (2001).

\bibitem{my} J. Korbicz, J.I. Cirac, J. Wehr, and M. Lewenstein, Phys. Rev. Lett. {\bf 94}, 153601 (2005).

\bibitem{glauber} E.C.G. Sudarshan, Phys. Rev. Lett. {\bf 10}, 277 (1963);
R.J. Glauber, Phys. Rev. {\bf 131}, 2766 (1963).

\bibitem{perelomov} A. Perelomov, {\it Generalized Coherent States and Their Applications} (Springer-Verlag, Berlin, 1986).

\bibitem{braunstein} S.L. Braunstein, C.M. Caves, R. Jozsa, N. Linden, S. Popescu, and R. Schack, Phys. Rev. Lett. {\bf 83}, 1054 (1999).

\bibitem{Kraus} B. Kraus, PhD Thesis, University of Innsbruck (2003).

\bibitem{vogel} Th. Richter and  W. Vogel, Phys. Rev. Lett. {\bf 89}, 283601 (2002).



\end{thebibliography}
\end{document}